# The Computational Power of
# Benenson Automata

David Soloveichik *

*Department of CNS, California Institute of Technology,
MC 136-93, 1200 E. California Blvd., 91125*

## Erik Winfree

*Department of Computer Science and CNS, California Institute of Technology,
MC 136-93, 1200 E. California Blvd., 91125*

**Abstract**

The development of autonomous molecular computers capable of making independent decisions *in vivo* regarding local drug administration may revolutionize medical science. Recently Benenson at el [3] have envisioned one form such a "smart drug" may take by implementing an *in vitro* scheme, in which a long DNA state molecule is cut repeatedly by a restriction enzyme in a manner dependent upon the presence of particular short DNA "rule molecules." To analyze the potential of their scheme in terms of the kinds of computations it can perform, we study an abstraction assuming that a certain class of restriction enzymes is available and reactions occur without error. We also discuss how our molecular algorithms could perform with known restriction enzymes. By exhibiting a way to simulate arbitrary circuits, we show that these "Benenson automata" are capable of computing arbitrary Boolean functions. Further, we show that they are able to compute efficiently exactly those functions computable by log-depth circuits. Computationally, we formalize a new variant of limited width branching programs with a molecular implementation.

## 1 Introduction

The goal of creating a molecular "smart drug" capable of making independent decisions *in vivo* regarding local drug administration has excited many researchers [6]. Recently, Benenson et al [3] (based on [4,2]) have envisioned what such an automaton may look like, and reported a partial implementation of the design *in vitro*. They made a system of an enzyme and a set of DNA molecules which tests whether particular RNA molecules are present in high concentration and other particular RNA molecules are present in low concentrations, and releases an output DNA molecule in high concentration only if the condition is met. The actual computation process consists of the enzyme cutting a special DNA molecule in a manner ultimately determined by the concentrations of input mRNA molecules present in solution. The authors suggest that such a design, or a similar one, can be used to detect concentrations of specific mRNA transcripts that are indicative of cancer or other diseases, and that the output can take the form of a "therapeutic" ssDNA.

---
* Corresponding author. Tel.: +001 626-395-6246
  *Email addresses:* `dsolov@caltech.edu` (David Soloveichik),
`winfree@caltech.edu` (Erik Winfree).

The key computational element in the scheme is an enzyme that cuts DNA in a controlled manner. Nature provides many biologically realizable methods of cutting DNA that can be adapted for computing. For instance, bacteria have evolved methods to cut the DNA of invading viruses (phages) with numerous enzymes called restriction enzymes. Most restriction enzymes cut double stranded DNA exclusively at sites where a specific sequence, called the recognition site, is found. Some restriction enzymes leave a so-called "sticky end overhang" which is a region of single stranded DNA at the end of a double stranded DNA molecule. Sticky ends are important because if there is another DNA molecule with a complementary sticky end, the two molecules can bind to each other forming a longer double stranded DNA strand.

Benenson et al use type IIS restriction enzymes, which cut double stranded DNA at a specific distance away from their recognition sites in a particular direction [7]. These enzymes were first considered in molecular computation by Rothemund [5] in an non-autonomous simulation of a Turing machine. For an example of a type IIS restriction enzyme, consider *Fok*I which is known to cut in the manner shown in Fig. 1(a). Note that after *Fok*I cuts, the DNA molecule is left with a sticky end overhang of 4 bases. The automaton of Benenson et al is based on a series of restriction enzyme cuts of a long *state molecule*. Each cut is initiated by the binding of a *cutting rule molecule* to the state molecule via matching sticky ends (Fig. 1(b)). Cutting rule molecules have an embedded restriction enzyme recognition site at a certain distance from their sticky end. The number of base pairs between the restriction enzyme recognition site and the sticky end on the cut rule molecule determines the number of bases that are cut away from the state molecule after the rule molecule attaches. Since the sequence of the sticky end on the state molecule determines which rule molecule attaches, it determines how many bases are cut off the state molecule in the presence of some set of rule molecules. Fig. 1(b) illustrates how TGGC can encode the "cut 7 bases" operation when the appropriate cutting rule molecule is present. After each cut, a new sticky end is revealed which encodes the location of the next cut, and the process can continue.

a)

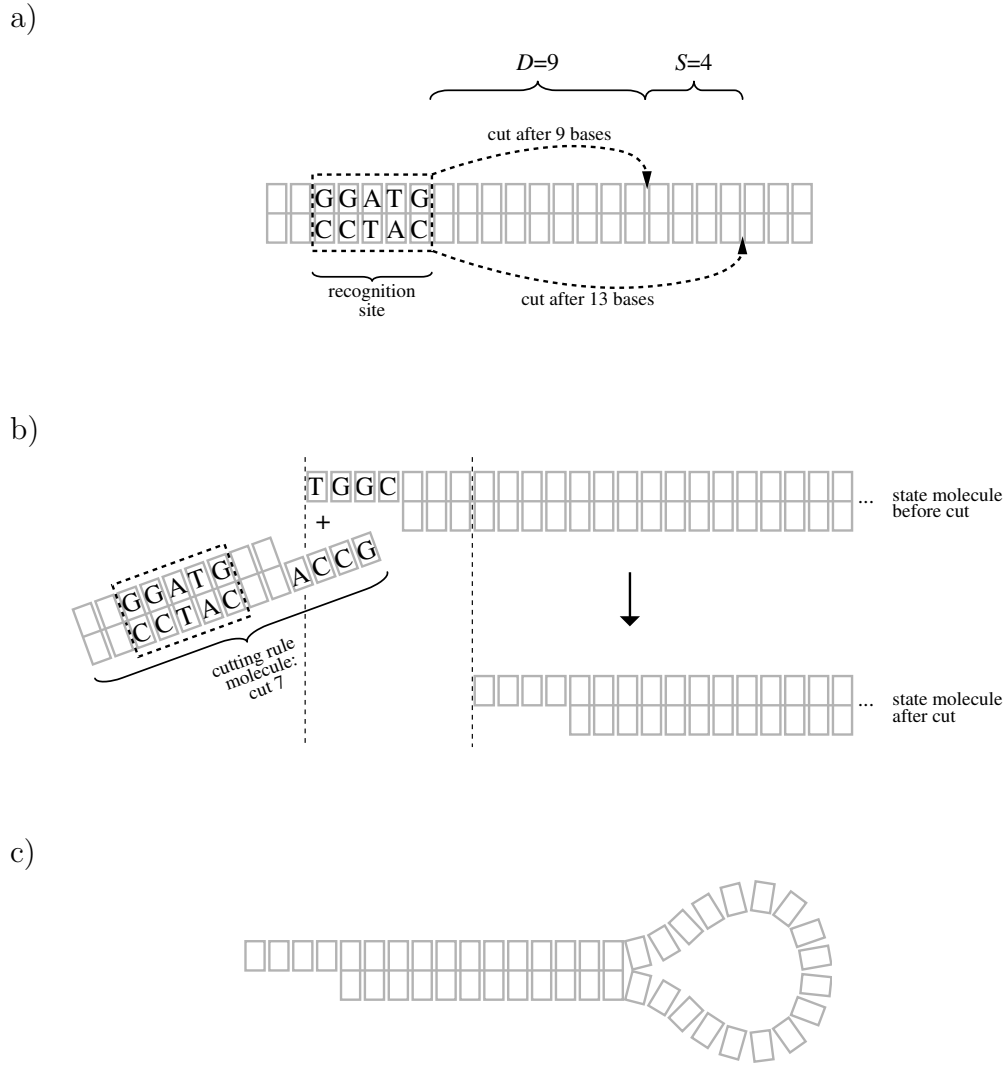

b)

c)

Fig. 1. (a) *Fok*I recognition and cut sites on a generic DNA substrate. The parameters $D$ and $S$ will be used to characterize restriction enzymes in this paper. $D$ is called the cutting range and $S$ the sticky end size. (b) Example of a cutting rule application. (c) Illustration of the output loop. Cutting beyond a certain point opens the loop. (In (a),(b),(c) the top strand is 5'→3'.)

Benenson et al [3] describe how any set of RNA or DNA molecules can act as input to their automaton. In particular, each input species converts some rule molecules that are initially inactive into active form, and inactivates others that are initially active. In this way, the presence of input molecules in either high or low concentration determines which rule molecules will be available.

If the state DNA molecule is cut beyond a certain point, a loop at the end is opened and released (Fig. 1(c)). Assuming the loop is inactive when closed and active when open, this results in the creation of the "theraputic" molecule in an input-dependent manner. If the system worked without error, and supposing that the input RNA molecules are either present in high concentration or not at all, the output DNA molecule should be released if and only if a set of RNAs is present that results in a set of rule molecules that cut the state DNA molecule sufficiently far. To accommodate the possibility of error, which we ignore here, they implement two possible outputs that compete between each other, with the one produced in largest quantities "winning."

In the first part of this paper, we formalize the computational process implemented by Benenson et al [3]. While the output DNA molecule released by one Benenson automaton can act as an input for another, allowing feedforward circuits to be implemented without difficulty, we would like to study the computational power of a system with a single state molecule. Showing how to compute complex functions with a single Benenson automaton makes it clear how one can compute even more complex functions with many state molecules. As part of our abstraction, we are going to ignore concentration dependence and other analog operations such as those involving competition between various reactions, and will focus on a binary model in which a reaction will either occur or not. We treat the state molecule and the set of possible cutting rule molecules as a program specifying what computation is to be performed, while the presence or absence of certain rule molecules is considered to be the input. Each rule molecule can be either always present, never present, present iff the corresponding input RNA species is present, or present iff the corresponding input RNA species is absent. We'll say that a Benenson automaton outputs 1 if at some point exactly $p$ bases are cut off, where $p$ indicates the last possible place to cut in the molecule. Otherwise, if this point is never reached, we say it outputs a 0. [1]

Like circuits, Benenson automata are best studied as a non-uniform computing model. But while the computational power of circuits is well characterized, the computational power of Benenson automata has not been studied. For example, while it was shown [3] how a single Benenson automaton can compute a conjunction of inputs (and negated inputs), it was not clear how a single Benenson automaton can compute a disjunction of conjunctions. While [4] and [2] show how finite automata can be simulated by a similar scheme [2], a different input method is used. Here, we show that a Benenson automaton can simulate an arbitrary circuit, implying that it is capable of doing arbitrary

---

[1] A moment's reflection will convince the reader that other possible output conventions such as indicating 0 or 1 by cutting before or after some $p$, or cutting to exactly $q$ or exactly $p$, would introduce no fundamental differences.

[2] In contrast to [3], [4,2] treat the state molecule as an input string for a uniform computation, while the set of rule molecules is always the same and specifies the finite state machine computation to be performed. It is interesting to note the difference in the computational power of these two approaches. To implement a FSM with $K$ symbols and $N$ states, a type IIS restriction enzyme with cutting range $N$ and sticky end size $O(\log KN)$ is sufficient and probably necessary.

non-uniform computation.

Lastly we study the cost of implementing more complex computations (e.g. more complex diagnostic tests) using Benenson automata. While increasing the length of the state molecule is relatively easy and incurs approximately linear cost, increasing the size of the sticky ends or the range at which the restriction enzyme cuts requires discovering or creating new enzymes. Suppose $\{f_n\}$ (where $f_n : \{0,1\}^n \rightarrow \{0,1\}$) is family of functions. We show that $f_n$ can be computed by Benenson automata such that the size of the sticky ends grows only logarithmically with $n$ and the range of the enzyme cutting stays constant.

Taking all the parameters of the model, we will say that a Benenson automaton computes a family of functions $\{f_n\}$ *efficiently* if it uses $O(\log(n))$ size sticky ends and $O(1)$ enzyme cutting range, and $poly(n)$ length state molecule. We will prove that under this definition of efficiency, Benenson automata are equivalent to $O(\log(n))$ depth circuits. This is the strictest definition of efficiency allowing $poly(n)$ length state molecules since at least $\log n$ size sticky ends are required to "read" all the input bits. We'll show, however, that allowing logarithmic cutting range cannot significantly increase the computational power of Benenson automata.

Independent of the relevance of our formalization to biological computation, Benenson automata capture a model of string cutting with input-dependent cutting rules, and may be of interest as such.

## 2    Formalization of Benenson automata

We consider Benenson automata over a fixed alphabet $\Sigma$. For biological plausibility, one may want to consider $\Sigma = \{A, T, C, G\}$. To simplify our constructions, we will assume $|\Sigma| \geq 3$; however, if so desired, the reader should be able to convert our constructions to a binary alphabet without excessive difficulty.

A Benenson automaton is parameterized by three numbers. The parameter $S$ corresponds to the sticky end size, $D$ to the maximum cutting range of the restriction enzyme (see Fig. 1(a)), and $L$ to the length of the state molecule. The revealed sticky end $\omega$ and the value of an input bit $x_i$ determine where the next cut will be made by the two cutting rules: $(i, 0, \omega, d)$ and $(i, 1, \omega, d')$, where $d$ is the distance cut [3] in the case that $x_i = 0$ and $d'$ in the case $x_i = 1$.

---

[3] Some Benenson automata may pose problems for existing or future restriction enzymes. For example, $d = 1$ would require a single base adjacent to a nick to be cleaved of each strand, which may not be biochemically plausible. Such difficulties can be avoided by minor modifications of our constructions.

We'll use the notation $\sigma[j]$ $(0 \leq j \leq |\sigma|)$ to indicate the string that is left over after the first $j$ symbols of $\sigma$ are stripped off. Interpreted as a DNA state molecule, the first $S$ symbols represent the single stranded sticky end overhang.

**Definition 1** *A $(S, D, L)$-Benenson automaton with $n$ inputs consists of a state string $\sigma \in \Sigma^L$ and of a set of input-dependent cutting rules $\mathcal{R}$ of the form: $(i, b, \omega, d)$ where $1 \leq i \leq n$, $b \in \{0, 1\}$, $\omega \in \Sigma^S$, and $d$ is an integer from 1 to $D$.*

**Definition 2** *Given a Benenson automaton, for a binary input $x$ of length $n$ and integers $j, j'$ $(0 \leq j < j' \leq L)$, we write $\sigma[j] \rightarrow_x \sigma[j']$ if $\exists (i, x_i, \omega, j' - j) \in \mathcal{R}$ where $x_i$ is the $i^{th}$ bit of the input and $\omega$ is the initial $S$ symbol portion of $\sigma[j]$. Further, $\rightarrow_x^*$ is the reflexive transitive closure of $\rightarrow_x$.*

**Definition 3** *A Benenson automaton is* deterministic *if it is impossible for conflicting cutting rules to apply simultaneously: if $(i, b, \omega, d), (i', b', \omega, d') \in \mathcal{R}$ where $d \neq d'$ then $i = i'$ and $b \neq b'$.*

**Definition 4** *We say that a deterministic Benenson automaton* computes *a boolean function $f : \{0, 1\}^n \rightarrow \{0, 1\}$ if there exists $p$ $(0 \leq p \leq L)$ such that $\sigma \rightarrow_x^* \sigma[p]$ if and only if $f(x) = 1$.*

If a Benenson automaton computes $f$ at position $p$, its state string can always be shortened to be at most $p$ symbols long. In a biochemical implementation the state molecule can be shortened and the loop placed such that the cut at $p$ is the last possible cut that can be made assuming the restriction enzyme cannot cut into the single stranded DNA of the loop. [4]

**Definition 5** *A Benenson automaton is* $s$-sparse *if $\forall i$ there are at most $s$ sticky ends $\omega$ s.t. $(i, 0, \omega, d), (i, 1, \omega, d') \in \mathcal{R}$ where $d \neq d'$.*

An $s$-sparse automaton has at most $s$ sticky ends "reading" any given input bit. In a biochemical implementation of a deterministic sparse Benenson automaton in order to change the input it is enough to change at most $s$ rule molecules per changed bit.

---

## 3   Characterizing the Computational Power of Benenson automata

In Section 4 we show that to compute function families using Benenson automata, only logarithmic scaling of the restriction enzyme parameters is needed, no matter what the complexity of the function family is. Further, if the families of functions is computable by log-depth circuits [5], then a state string of only polynomial size is required:

**Theorem 6**

(a) *Any function $f : \{0,1\}^n \to \{0,1\}$ can be computed by a $(S, D, L)$-Benenson automaton where $S = O(\log n)$ and $D = O(1)$.*

(b) *Families of functions computable by $O(\log n)$ depth circuits can be computed by $(S, D, L)$-Benenson automata where $S = O(\log n)$, $D = O(1)$, and $L = poly(n)$.*

The constants implicit in both statements are rather small. Note that the sticky end size cannot be smaller than $O(\log n)$ since there must be at least a different sticky end for each input (otherwise the input isn't completely "read"). Thus, in computing arbitrary functions, we cannot do better than $S = O(\log n)$ and $D = O(1)$.

Further, in Section 5 we prove that our computation of families of functions computable by log-depth circuits is optimal:

**Theorem 7**  *Families of functions computable by $(O(\log n), O(1), poly(n))$-Benenson automata can be computed by $O(\log n)$-depth circuits.*

**Corollary 8**  *$(O(\log n), O(1), poly(n))$-Benenson automata can compute the same class of families of functions as $O(\log n)$-depth circuits.*

So if we consider only $(O(\log n), O(1), poly(n))$-Benenson automata efficient, then Benenson automata can compute a family of non-uniform functions efficiently if and only if it can be computed by a circuit of logarithmic depth. In Section 5, we'll also show that relaxing this notion of efficiency to include logarithmic cutting range does not increase the computational power significantly.

---

[5] For the purposes of this paper, circuits are feed-forward and consist of AND, OR, NOT gates with fan-in bounded by 2.

## 4 Simulating Branching Programs and Circuits

Benenson automata are closely related to the computational model known as branching programs. In the next section we show how arbitrary branching programs can be simulated. In the following two sections, we show how restricted classes of branching programs (fixed-width and permutation branching programs) can be simulated by Benenson automata with $S = O(\log n)$ and $D = O(1)$. Since fixed-width permutation branching programs are still powerful enough to compute arbitrary functions (Section 4.4), Theorem 6(a) follows. Further, in Section 4.4 we'll also see that fixed-width permutation branching programs of $poly(n)$ size can simulate $O(\log n)$ depth circuits, implying Theorem 7(b).

### 4.1 General Branching Programs

A branching program is a directed acyclic graph with three types of nodes: variable, accept and reject. (E.g. Fig. 2(a).) The variable nodes are labeled with an input variable $x_i$ $(1 \le i \le n)$ and have two outgoing edges, one labeled 0 and the other 1, that lead to other variable accept or reject nodes. The accept and reject nodes have no outgoing edges. One variable node with no incoming edges is designated the start node. The process of computation consists of starting at the start node and at every node $x_i$, following the outgoing edge whose label matches the value of the $i^{th}$ bit of the input. If an accept node is reached, we say that the branching program accepts the input $x$. Otherwise, a reject node is reached, and we say that the branching program rejects the input $x$. The function $f : \{0,1\}^n \to \{0,1\}$ computed by a branching program is $f(x) = 1$ if $x$ is accepted and 0 otherwise.

The naive way of simulating a branching program with a Benenson automaton is as follows. Because a branching program is a directed acyclic graph, we can index the nodes in such a way that we can never go from a node with a higher index to a node with a lower one (as shown in Fig. 2(a)). We can ensure that the first node is the start node and that there is only one accept node (we can convert all other accept nodes into variable nodes with all outgoing edges to this accept node). The state string consists of consecutive segments, with unique sequences of length $S$, for each node in the branching program. For each node $q$ the segment $\sigma_q$ is such that $\sigma_q \cdots \sigma_{q'} \cdots \to_x \sigma_{q'} \cdots$ iff the branching program goes to node $q'$ from $q$ on input $x$. This is implemented by providing two cutting rule molecules, one for each case. To be explicit, for every variable node $q$ labelled $x_i$, define $var(q) = i$. Further, $goto_0(q)$ is the node targeted by the 0 outgoing edge and $goto_1(q)$ is the node targeted by the 1 outgoing edge of $q$. Using this notation, we need the following cutting rules for every variable node $q$: $(var(q), 0, \sigma_q, (goto_0(q) - q)S)$ and $(var(q), 1, \sigma_q, (goto_1(q) - q)S)$. We cut to the beginning of the segment corresponding to the accept node iff the branching program accepts the input $x$. See Fig. 2(a,b) for an example of a branching program and the corresponding Benenson automaton. To implement the automaton as described, we need $S = \lceil \log_{|\Sigma|}(H) \rceil$ where $H$ is the number of nodes in the branching program. The maximum cutting range $D$ may need to be $(H-1)S$ if in our branching program we may jump from the first node to the accept node. The size of the state string $L$ needs to be $HS$. Thus we have the following lemma:

**Lemma 9** *A function computed by a branching program of $H$ nodes can be computed by a $(S, D, L)$-Benenson automaton where $S = \lceil \log_{|\Sigma|}(H) \rceil$, $D = (H-1)S$, and $L = HS$.*

Note that all three complexity parameters ($S$, $D$, and $L$) of Benenson automata needed to simulate general branching programs using the above construction scale with the size of the branching program. Thus, for families of functions for which the size of branching programs computing them increases very fast with $n$, new restriction enzymes must be developed that scale similarly. Thus this is not enough to prove Theorem 6.

### 4.2 Fixed-Width Branching Programs

In this section, we demonstrate a sufficiently powerful subclass of branching programs whose simulation is possible by Benenson automata such that only the size of the state molecule ($L$) scales with the the size of the branching program, while $S = O(\log n)$ and $D = O(1)$.

In the general case discussed in Sec. 4.1, our cutting range had to be large because we had no restriction on the connectivity of the branching program and may needed to jump far. Further, we used a different sticky end for each node because there may be many different "connectivity patterns." Restricting the connectivity of a branching program in a particular way permits optimizing the construction to significantly decrease $S$ and $D$. In fact, both will loose their dependence on the size of the branching program. In the final construction, the sticky end size $S$ will depend only on the number of variables and the cutting range will be a constant.

A width $J$, length $K$ branching program consists of $K$ layers of $J$ nodes each. (E.g. Fig. 2(c).) We will think of $J$ as a constant since for our purposes $J \leq 5$ will be enough. (The total number of nodes is of course $H = KJ$.) Nodes in each layer have outgoing edges only to the next layer, and every node in the last layer is either accepting or rejecting. We can ensure that the first node in the first layer is the start node and that the last layer has a single accept

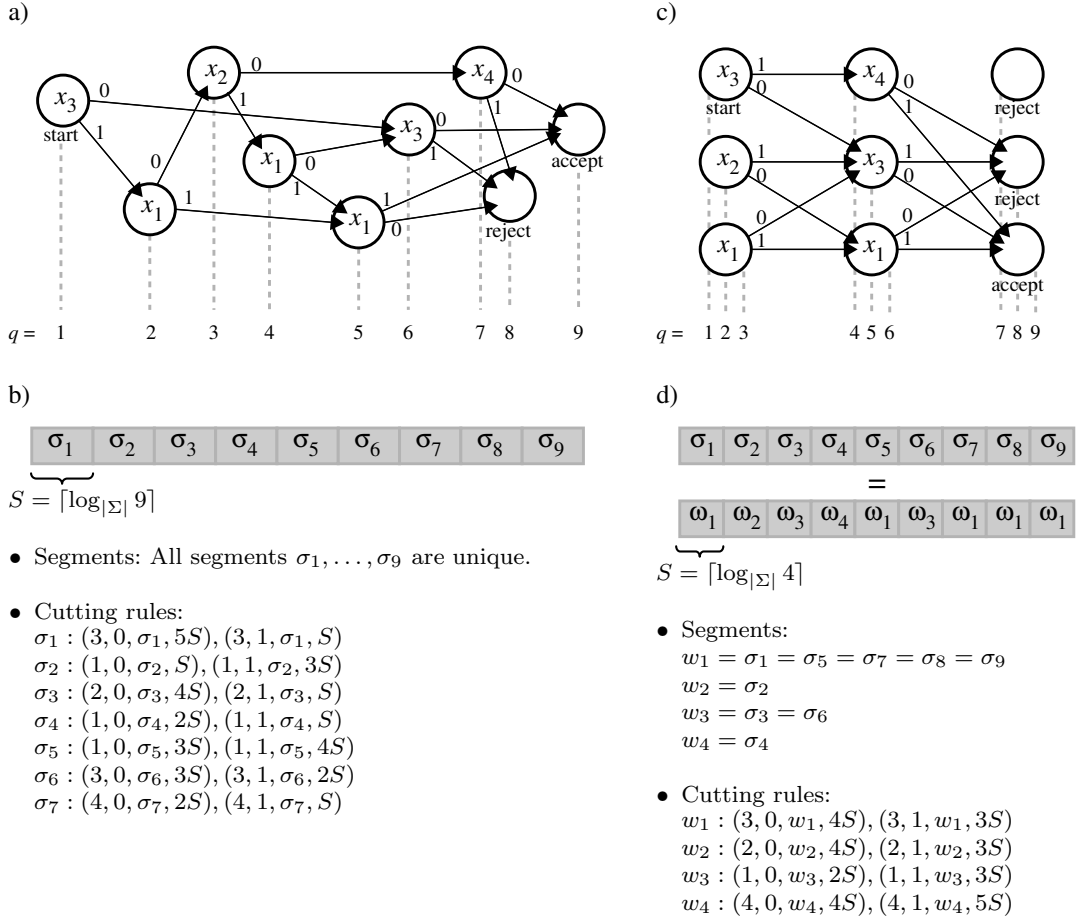

a)

b)

| $\sigma_1$ | $\sigma_2$ | $\sigma_3$ | $\sigma_4$ | $\sigma_5$ | $\sigma_6$ | $\sigma_7$ | $\sigma_8$ | $\sigma_9$ |

$$\underbrace{\phantom{\sigma_1}}\;S = \lceil \log_{|\Sigma|} 9 \rceil$$

- Segments: All segments $\sigma_1, \ldots, \sigma_9$ are unique.

- Cutting rules:
  $\sigma_1 : (3, 0, \sigma_1, 5S), (3, 1, \sigma_1, S)$
  $\sigma_2 : (1, 0, \sigma_2, S), (1, 1, \sigma_2, 3S)$
  $\sigma_3 : (2, 0, \sigma_3, 4S), (2, 1, \sigma_3, S)$
  $\sigma_4 : (1, 0, \sigma_4, 2S), (1, 1, \sigma_4, S)$
  $\sigma_5 : (1, 0, \sigma_5, 3S), (1, 1, \sigma_5, 4S)$
  $\sigma_6 : (3, 0, \sigma_6, 3S), (3, 1, \sigma_6, 2S)$
  $\sigma_7 : (4, 0, \sigma_7, 2S), (4, 1, \sigma_7, S)$

c)

d)

| $\sigma_1$ | $\sigma_2$ | $\sigma_3$ | $\sigma_4$ | $\sigma_5$ | $\sigma_6$ | $\sigma_7$ | $\sigma_8$ | $\sigma_9$ |

$$=$$

| $\omega_1$ | $\omega_2$ | $\omega_3$ | $\omega_4$ | $\omega_1$ | $\omega_3$ | $\omega_1$ | $\omega_1$ | $\omega_1$ |

$$\underbrace{\phantom{\omega_1}}\;S = \lceil \log_{|\Sigma|} 4 \rceil$$

- Segments:
  $w_1 = \sigma_1 = \sigma_5 = \sigma_7 = \sigma_8 = \sigma_9$
  $w_2 = \sigma_2$
  $w_3 = \sigma_3 = \sigma_6$
  $w_4 = \sigma_4$

- Cutting rules:
  $w_1 : (3, 0, w_1, 4S), (3, 1, w_1, 3S)$
  $w_2 : (2, 0, w_2, 4S), (2, 1, w_2, 3S)$
  $w_3 : (1, 0, w_3, 2S), (1, 1, w_3, 3S)$
  $w_4 : (4, 0, w_4, 4S), (4, 1, w_4, 5S)$

Fig. 2. (a) An example of a general branching program of 9 nodes over 4 inputs and (b) the corresponding Benenson automaton. (c) An example of a width 3 branching program of 9 nodes over 4 inputs and (d) the corresponding Benenson automaton. Note that some nodes are inaccessible but these will be a small fraction for large programs. In both examples, $\sigma_1 \cdots \sigma_9 \rightarrow_x^* \sigma_9$ iff the branching program accepts $x$.

node. (Otherwise, the branching program can be trivially modified.) It turns out that width 5 branching programs are sufficiently powerful to simulate any circuit (Section 4.4). Further, the results of Section 5 ensure that we have not restricted our model of computation too much; more general Benenson automata can not compute more efficiently.

To simulate width $J$ branching programs using the above method we index nodes consecutively from each layer: the $j$th node in layer $k$ obtains index $q = (k-1)J+j$. We need the maximum cutting range to be only $D = (2J-1)S$ since in the worst case we need to go from the first node of a layer to the last node of the next layer. Further, we don't need a unique segment for each node in the branching program. We just need a unique segment for each $(i, \Delta q_0, \Delta q_1)$ combination used, where $\Delta q_0 = goto_0(q) - q$ is the number of segments to skip in the case $x_i = 0$ and $\Delta q_1 = goto_1(q) - q$ is the number of segments to skip in the case $x_i = 1$. For a width $J$ branching program, $\Delta q_0$ and $\Delta q_1$ range from 1 to $2J - 1$. So at most we need $S = \lceil \log_\Sigma(n(2J-1)^2) \rceil$, and the resultant automaton is $(2J-1)^2$-sparse. (The segments corresponding to the accept and reject nodes can be anything as long as we cannot go from a reject node to the accept node. We can choose a segment such that $\Delta q_0, \Delta q_1 \geq J$.) Note that $S$ and $D$ loose their dependence on the length of the branching program $K$. This means that if the width $J$ is fixed, then the restriction enzyme needed is independent of the size of the branching program and is dependent only on the length of the input $n$. See Fig. 2(c,d) for an example of how a fixed-width branching program can be converted to a Benenson automaton.

**Lemma 10** *A function over $n$ inputs computed by a branching program of width $J$ and length $K$ can be computed by a $(2J-1)^2$-sparse $(S, D, L)$-Benenson automaton where $S = \lceil \log_\Sigma(n(2J-1)^2) \rceil$, $D = (2J-1)S$, and $L = KJS$.*

The constructions described above rely on being able to skip entire segments in a single cut. It seems that the cutting range must be at least logarithmic in $n$, since the size of the segments is logarithmic in $n$ to be able to read all the input variables. However, surprisingly, for fixed-width branching programs we can attain constant cutting range $D$ dependent only on the width $J$ and not on $n$ or the number of layers $K$. The segments can be designed in such a way that cutting $d$ symbols $(1 \leq d \leq D)$ from the beginning of any segment $\sigma_q$ is followed by consecutive applications of special input-independent *skip* cutting rules until $d$ entire segments have been cut off, at which point the input-dependent *segment* cutting rule specified by the segment $\sigma_{q+d}$ becomes applicable. In other words, $\sigma_q[d]\sigma_{q+1}\sigma_{q+2}\cdots \rightarrow_x^* \sigma_{q+d}\sigma_{q+d+1}\cdots$ for all possible sequences of segments. This method allows the segments (which will be $O(\log n)$) to be much longer than the cutting range ($O(1)$) and yet be able to specify an input-dependent jump to one of $D$ following segments. Fig. 3 shows an example of a segment cutting rule application that cuts 2 symbols, followed by applications of skip cutting rules. Specifically, let's assume $|\Sigma| \geq 3$ and let $\iota \in \Sigma$. Symbol $\iota$ marks the beginning of each segment $\sigma_q$ and the rest of the segment uses symbols in $\Sigma - \{\iota\}$. Segments are of length $m = D \cdot k + 1$ for some integer $k \geq 1$. Note that the segments are longer than the sticky ends. Only the initial $S$ symbols of the segment will be "coding" (gray squares in Fig. 3). The input-dependent segment cutting rules are only applicable if the first symbol of the revealed sticky end is $\iota$. The input-independent skip cutting rules say that if the first symbol of the revealed sticky end is not $\iota$ then cut $D$ symbols. [6] So every time the boundary of a segment is crossed by the skip cutting rules, the misalignment created by the initial cut of $\sigma_q$ by the segment cutting rule decreases by one symbol, until the cut is fully aligned with the beginning of the segment $\sigma_{q+d}$ allowing the application of the corresponding segment cutting rule.

---

[6] Formally, an input-independent cutting rule $(\cdot, \cdot)$ is a pair of normal (input-dependent) cutting rules $(1, 0, \cdot, \cdot)$ and $(1, 1, \cdot, \cdot)$. Thus, the skip cutting rules are all rules of the form: $(\tau, D)$ where $\tau \in \Sigma^S$ and the first symbol of $\tau$ is not $\iota$.

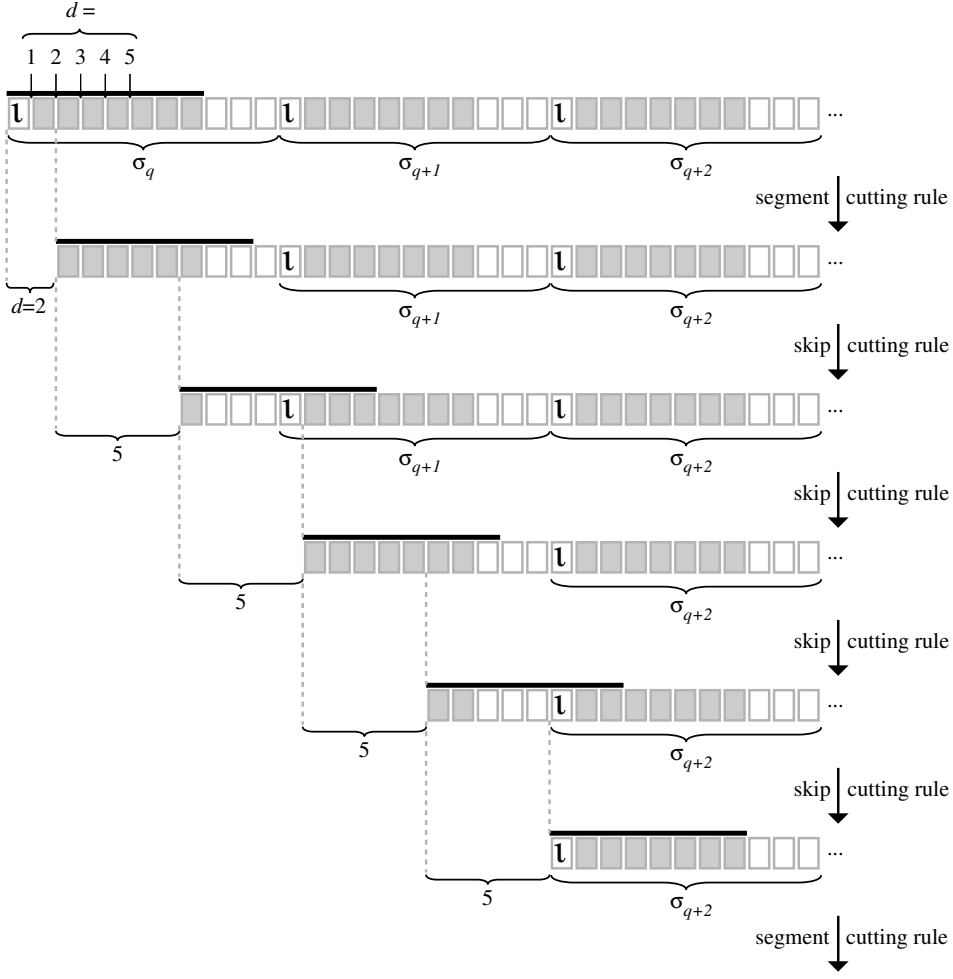

Fig. 3. An example of a segment cutting rule application and the subsequent application of skip cutting rules. In this case, $D = 5$, $k = 2$ and the size of the segments is $m = D \cdot k + 1 = 11$. The sticky end size is $S = 8$; the black horizontal lines above the state string show the sticky end in each step. The grayed squares indicate the coding symbols that, together with a bit of input, determine which segment cutting rule is applicable.

With the above trick, we have the following lemma for fixed-width branching programs:

**Lemma 11** *A function over $n$ inputs computed by a branching program of width $J$ and length $K$ can be computed by a $(2J-1)^2$-sparse $(S, D, L)$-Benenson automaton where $S = 1 + \lceil \log_{|\Sigma|-1} (n(2J-1)^2) \rceil$, $D = 2J-1$, and $L = KJS$.*

Lemma 11 together with Barrington's theorem (Lemma 13) is enough to prove both parts of Theorem 6. However, we first optimize our construction even further to obtain better constants.

## 4.3 Permutation Branching Programs

We can attain better constants if we restrict the branching program even more. Again, in the next section we'll see that even with this restriction, branching programs can simulate circuits.

First, we need a notation for the context of layered branching programs. For node $j$ in layer $k$ let $goto_0(k, j) = j'$ if the $j'$th node in layer $k + 1$ is targeted by the 0 outgoing edge of this node; $goto_1(k, j)$ is defined similarly. A width $J$ permutation branching program is a width $J$ branching program such that for all $k$, $goto_0(k, \cdot)$ and $goto_1(k, \cdot)$ are permutations. Further, there is exactly one accept node in the last layer (this can no longer be trivially assumed). It turns out that width 5 permutation branching programs are still sufficiently powerful to simulate any circuit (Section 4.4). In Section 5, we'll confirm that we have not restricted our model of computation too much: efficient Benenson automata cannot simulate anything more powerful than permutation branching programs.

It is easy to see that for every permutation branching program, there is another permutation branching program of the same width and length that accepts the same inputs as the original program but for all $k$, $goto_0(k, \cdot)$ is the identity permutation (i.e. $goto_0(k, j) = j$). In this case, we only need a unique segment for every $(i, \Delta q_1)$ combination, since $\Delta q_0$ is always $J$. This leads to the following lemma:

**Lemma 12** *A function over $n$ inputs computed by a permutation branching program of width $J$ and length $K$ can be computed by a $(2J - 1)$-sparse $(S, D, L)$-Benenson automaton where $S = 1 + \lceil \log_{|\Sigma|-1} (n(2J-1)) \rceil$, $D = 2J - 1$, and $L = KJS$.*

## 4.4 Simulating Circuits

While it may seem that fixed-width permutation branching programs are a very weak model of computation, it turns out that to simulate circuits, width 5 permutation branching programs is all we need:

**Lemma 13 (Barrington [1])** *A function over $n$ variables computed by a circuit of depth $C$ can be computed by a length $4^C$ width 5 permutation branching program.*

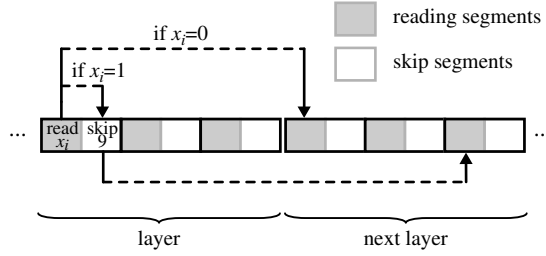

Fig. 4. Illustration of the construction achieving 1-sparseness. In this case the width of the branching program is $J = 3$. Note that each skip illustrated by the dashed lines consists of many cuts like those illustrated in Fig. 3.

**Corollary 14 (of Lemmas 12 and 13)** *A function over $n$ variables computed by a circuit of depth $C$ can be computed by a 9-sparse $(S, D, L)$-Benenson automaton with $S = 1 + \lceil \log_{|\Sigma|-1}(9n) \rceil$, $D = 9$, and $L = 4^C 5S$.*

This provides an alternate an alternative proof of Theorem 6 and implies, for instance, that a Benenson automaton using *Fok*I can do arbitrary 3-bit computation. Any increase in the sticky end size, exponentially increases the number of inputs that can be handled. If an enzyme is discovered that cuts 9 bases away like *Fok*I but leaves size 7 sticky ends, then it can do all 81-bit computation.

Letting $C = O(\log n)$, this proves Theorem 6(b). Theorem 6(a), of course, follows trivially since the complexity of the circuit ($C$) enters only in the length of the state string.

*4.5   1-sparseness*

If 1-sparseness is essential, the above scheme can be adapted at the expense of slightly increasing the maximum cutting range $D$ and the length of the state string $L$. The sticky end size can be actually decreased. Instead of a unique segment for each $(i, \Delta q_1)$ pair, we can use a pair of segments for each node of the permutation branching program (see Fig. 4). The first segment (the *reading* segment) reads the corresponding variable and skips either $2J$ segments if $x_i = 0$ or goes to the next segment if $x_i = 1$. Then the next segment (the *skip* segment) encodes an input-independent skip of $2\Delta q_1 - 1$ segments to go to the correct reading segment. We need at most $n + 2J - 1$ unique segment types: $n$ to read all the variables, and $2J - 1$ to be able to skip $2\Delta q_1 - 1$ segments for each $1 \leq \Delta q_1 \leq 2J - 1$. The maximum number of segments to skip is $2(2J - 1) - 1 = 4J - 3$.

**Lemma 15** *A function over $n$ inputs computed by a permutation branching program of width $J$ and length $K$ can be computed by a 1-sparse $(S, D, L)$-Benenson automaton where $S = 1 + \lceil \log_{|\Sigma|-1}(n + 2J - 1) \rceil$, $D = 4J - 3$, and $L = 2KJS$.*

This implies, for instance, that 1-sparse Benenson automata using *Fok*I can simulate any width 3 permutation branching program over 22 inputs.

**Corollary 16 (of Lemmas 15 and 13)** *A function over $n$ variables computed by a circuit of depth $C$ can be computed by a 1-sparse $(S, D, L)$-Benenson automaton with $S = 1 + \lceil \log_{|\Sigma|-1}(n + 9) \rceil$, $D = 17$, and $L = 4^C 30S$.*

This implies, for example, that if a DNA restriction enzyme can be found that leaves sticky ends of size 4 like *Fok*I but cuts 17 bases away, then this enzyme can do all 18 bit computation with 1-sparse Benenson automata.

## 5 Shallow Circuits to Simulate Benenson automata

Function families computable by Benenson automata with $S = O(\log n)$, $D = O(1)$, and $L = poly(n)$ can be computed by $\log n$ depth circuits, proving Theorem 7.

**Lemma 17** *A function computed by a $(S, D, L)$-Benenson automaton can be computed by a $O(\log(L/D) \log D + D)$ depth, $O((\log D) 2^D L)$ size circuit.*

To see that this Lemma implies Theorem 7, take $D = O(1)$, $S = O(\log n)$, and $L = poly(n)$. Note that since width 5 permutation branching programs are equivalent to log-depth circuits [1], this lemma also establishes that efficient Benenson automata are not more powerful than width 5 permutation branching programs.

Lemma 17 also implies that $(O(\log n), O(\log n), poly(n))$-Benenson automata cannot be much more powerful than $(O(\log n), O(1), poly(n))$-Benenson automata. Specifically, $\forall \varepsilon > 0$, functions computable by $(O(\log n), O(\log n), poly(n))$-Benenson automata are computable by $O(\log^{1+\varepsilon} n)$ depth, $poly(n)$ size circuits. Interestingly, note that sticky end size $S$ does not affect the complexity of the circuit simulating a Benenson automaton.

The idea of our construction is that we split the state string into segments of length $D$ and compute for all possible cuts in each segment, for the given input, where the last cut in the next segment would be. Then a binary tree of compositions of these functions results in a function from which it is easy to determine whether the state string starting from the beginning will be cut to the point that indicates that the output of the Benenson automaton is 1.

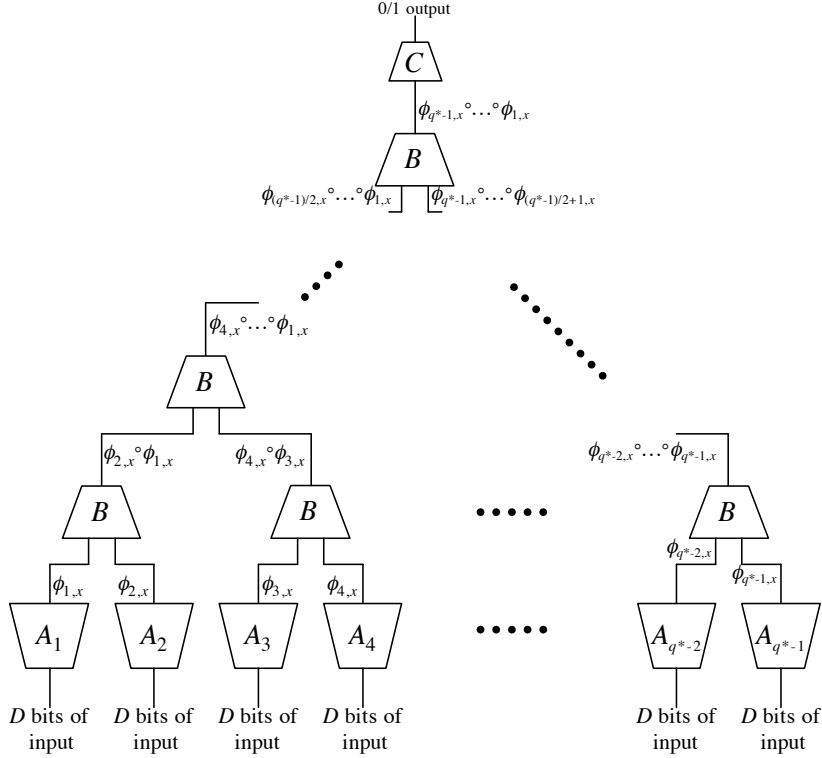

Fig. 5. Circuit outline for simulating a Benenson automaton. The $\phi$ lines represent a bundle of $D\lceil \log(D+1)\rceil$ wires. Input lines represent a bundle of at most $D$ wires (a different subset for each gadget, possibly overlapping).

Note that the segments cannot be shorter than $D$ since then a cut may entirely bypass a segment, and making them longer than $D$ makes the construction less efficient (i.e. results in a deeper circuit.) This proof is similar to the proof that poly-length fixed-width branching programs can be simulated by log-depth circuits (e.g. [1]), in which composition circuits are computing compositions of permutations.

Suppose $\sigma \to_x^* \sigma[p] \Leftrightarrow f(x) = 1$. We split the state string $\sigma$ into segments of length $D$ (except for the last segment which may contain fewer than $D$ symbols): $\sigma = \sigma_1, \ldots, \sigma_{\lceil L/D\rceil}$. To each segment $\sigma_q$, there corresponds a function $\phi_{q,x} : \{0, \ldots, D-1, \perp\} \to \{0, \ldots, D-1, \perp\}$. For each possible cut in $\sigma_q$, $\phi_{q,x}$ returns the location of the last cut in the next segment, if such can be made. Formally, $\phi_{q,x}(j)$ returns the largest $h$ ($0 \leq h \leq D-1$), if such exists, s.t. $\sigma_q[j]\sigma_{q+1} \cdots \sigma_{\lceil L/D\rceil} \to_x^* \sigma_{q+1}[h]\sigma_{q+2}\cdots\sigma_{\lceil L/D\rceil}$. Otherwise, $\phi_{q,x}$ returns $\perp$. Further, $\phi_{q,x}(\perp) = \perp$. Let $\sigma_{q^*}$ be the segment containing the cut point that detects the output of the Benenson automaton. In other words, $\sigma_{q^*}[j^*]\sigma_{q^*+1}\cdots\sigma_{\lceil L/D\rceil} = \sigma[p]$ for $0 \leq j^* \leq D-1$. Then, $\phi_{q^*-1,x} \circ \cdots \circ \phi_{2,x} \circ \phi_{1,x}(0) = j^* \Leftrightarrow f(x) = 1$.

For clarity of exposition, let us assume that $q^* - 1$ is a power of 2. Our circuit consists of gadgets $A_q$ ($1 \leq q \leq q^*-1$), gadgets $B$ and gadget $C$ (see Fig. 5).

The input and output lines of gadgets $B$ represent some composition $\phi$ of $\phi_{q,x}$'s by a table of $D$ rows of $\lceil \log{(D+1)} \rceil$ bits each such that the $j$th row (0 indexed) has value $\phi(j)$. To compute the initial series of tables, each gadget $A_q$ needs only to know at most $D$ bits of input $x$ on which $\phi_{q,x}$ may depend (a segment of length $D$ can read at most $D$ input bits). Each gadget $A_q$ can be a selector circuit that uses the relevant input bits to select one of $2^D$ possible hardwired outputs (different for each $q$). These gadgets $A_q$ have depth $O(D)$ and size $O(D \log{(D)} 2^D)$. All gadget $B$ needs to do to compute the $j$th row of its output is to look at the $j$th row of its first input (say it contains number $h$) and produce the $h$th row of its second input, or if the $D$th row of its first input contains $\perp$ then produce $\perp$. Gadget $B$ is just a selector circuit and has depth $O(\log D)$ and size $O(D^2 \log D)$. Gadget $C$ needs to check if the 0th row of its input table contains the number $j^*$ and output 1 if so and 0 otherwise.

## 6  Discussion

This work generalizes the non-uniform model of computation based on the work of Benenson et al [3] and characterizes its computational power. We considered restriction enzymes with variable reach and sticky end size, and studied how the complexity of the possible computation scales with these parameters. In particular, we showed that Benenson automata can simulate arbitrary circuits and that polynomial length Benenson automata with constant cutting range are equivalent to fixed-width branching programs and therefore equivalent to log-depth circuits. We achieve these asymptotic results with good constants suggesting that the insights and constructions developed here may have applications.

There may be ways to reduce the constants in our results even further. Although the fixed-width permutation branching programs produced via Barrington's theorem have the same variable read by every node in a layer, this fact is not used in our constructions. Exploiting it may achieve smaller sticky end size or maximum cutting distance.

As mentioned in Section 2, in a biochemical implementation of our constructions the last possible cut in the case that $f(x) = 0$ may have to be sufficiently far away from the output loop to prevent its erroneous opening. By using a few extra unique sticky ends we can always achieve this. For example, by adding one more unique sticky end corresponding to the reject states and making sure the accept state is last, we can ensure that in the constructions simulating general branching programs and fixed-width branching programs the last possible cut in the case $f(x) = 0$ is at least the length of a segment away ($\geq S, D$) from the last cut in the case $f(x) = 1$.

Nevertheless, the major problem with directly implementing our construction is the potential of error during the attachment of the rule molecule and during cuts. While a practical implementation of a Benenson automaton [3] has to work reliably despite high error rates, our formalization does not take the possibility of erroneous cutting into account. Further work is needed to formalize and characterize effective error-robust computation with Benenson automata. Similarly, while it is easiest to study the binary model in which a reaction either occurs or not, a model of analog concentration comparisons may better match some types of tests implemented by Benenson et al.